\documentclass[aps,prd, twocolumn,floatfix,showpacs]{revtex4}
\usepackage{graphicx}
\usepackage{amssymb}

\begin{document}

\title{{\large \bf Observational constraints on braneworld geometry}}

\author{Gonzalo A. Palma}

\email{G.A.Palma@damtp.cam.ac.uk}

\affiliation{\mbox{Department of Applied Mathematics and Theoretical
Physics, Centre for Mathematical Sciences,} University of Cambridge,
Wilberforce Road, Cambridge CB3 0WA, United Kingdom}

\date{February 2006}

\begin{abstract}
The low energy regime of 5D braneworld models with a bulk scalar
field is studied. The setup is rather general and includes the
Randall-Sundrum and dilatonic braneworlds models as particular
cases. We discuss the cosmological evolution of the system and
conclude that, in a two brane system, the negative tension brane is
generally expected to evolve towards a null warp-factor state. This
implies, for late time cosmology, that both branes end up
interacting weakly. We also analyze the observational constraints
imposed by solar-system and binary-pulsar tests on the braneworld
configuration. This is done by considering the small deviations
produced by the branes on the 4D gravitational interaction between
bodies in the same brane. Using these constraints we show that the
geometry around the braneworld is strongly warped, and that both
branes must be far apart.
\end{abstract}


\maketitle

%----------------------------------------------------------------------------Introduction

\section{Introduction} \label{sec: Intro}

Braneworld models have been shown to be extremely rich in phenomena
leading to modifications of General Relativity (GR) at both low and
high energies \cite{Langlois, Maartens, brax-bruck-davis}. When the
four-dimensional description of branes is considered, massless
scalar degrees of freedom -the moduli- appear mediating the
gravitational interaction together with the usual 4D graviton. The
presence of these moduli is commonly related to the geometry of the
entire system and plays a significant role in the phenomenology of
extra-dimensions. At low energies, for instance, braneworld models
are best described by scalar-tensor theories \cite{scalar-tensor1,
scalar-tensor2, scalar-tensor3, scalar-tensor4, scalar-tensor5,
Kanti1, Kanti2, Palma-Davis1, Palma-Davis2}. Within this framework,
the gravitational couplings become functions of the moduli and
standard GR predictions get modified in ways that can be tightly
constrained by present astrophysical observations.

Current tests on gravity allow us to measure relativistic
corrections to Newton's law at the first post-Newtonian level
($\propto 1/c^{2}$) with great precision \cite{Edd}. At this order,
the predictions of scalar-tensor theories can be parameterized by
two ``weak field'' quantities, $\beta$ and $\gamma$, the
post-Newtonian parameters introduced long ago by Eddington
\cite{Eddington}. When solar-system tests are taken into account
these parameters are observed to be very close to 1 (GR corresponds
to $\beta = \gamma=1$). For example, the time delay variation of the
Cassini spacecraft near the solar conjunction \cite{Cassini} has
given the result $\gamma - 1 = (2.1 \pm 2.3) \times 10^{-5}$,
whereas the lunar laser ranging experiment \cite{Lunar} shows that
$4\beta - \gamma - 3 = (-0.7 \pm 1) \times 10^{-3}$. Another source
of constraints on $\beta$ and $\gamma$, qualitatively different from
solar-system tests, comes from the observation on binary pulsars
systems \cite{Pulsar1, Pulsar2}. In this case, the ``strong field''
conditions generated by the compactness of neutron stars allow the
analysis of nonperturbative effects caused by the moduli, implying
tight bounds on the Eddington parameters.

In this paper the low energy regime of braneworld models is
considered. We show that solar-system and binary-pulsar tests can be
used to discriminate between different 5D braneworld models and to
shed light on their extra-dimensional configuration. For this, we
have chosen a rather general setup, BPS-braneworlds, of which the
Randall-Sundrum model \cite{RS1, RS2} and dilatonic braneworlds
\cite{Lukas} are just particular examples. The model consists of a
5D space-time bounded by two 3-branes with tensions $\lambda_{1}$
and $\lambda_{2}$. A scalar field $\phi$ lives in the 5D bulk and is
dominated by a bulk potential $U(\phi)$. At the same time, the brane
tensions are functions of the scalar field boundary values,
$\phi^{1}$ and $\phi^{2}$, of the form: $\lambda_{1} = +
\sigma(\phi^{1})$ and $\lambda_{2} = - \sigma(\phi^{2})$, where
$\sigma(\phi)$ is the brane potential. In order to stay close to GR
and acceptable phenomenology \cite{pheno1, pheno2}, a special
condition (the BPS condition) exists between the scalar field
potential $U(\phi)$ and the brane potential $\sigma(\phi)$:
\begin{eqnarray} \label{S1: cond}
U = (\partial_{\phi} \sigma)^{2} - \sigma^{2}.
\end{eqnarray}
This condition ensures supersymmetry to be preserved near the branes
\cite{SUSY1, SUSY2, SUSY3, SUSY4, SUSY5} and the existence of static
vacuum solutions in the absence of matter fields. Additionally, it
allows the system to be described by a simple and attractive
bi-scalar-tensor theory, where the moduli ($\phi^{1}$ and
$\phi^{2}$) are related to the positions of the branes on the 5D
background. Deviations from the BPS condition can also be included
in the formalism, resulting in the presence of dark energy.

This article is organized as follows: In Sec. \ref{S2} we review BPS
braneworlds and introduce their low energy effective description. We
also discuss some relevant aspects of these models, such as its
extra-dimensional geometry and the presence of singularities. In
Sec. \ref{S3} we deduce the system's equations of motion and analyze
the cosmological evolution of the branes. There we find that the
negative tension brane is generally expected to evolve towards a
state where its warp factor becomes null. For late time cosmology
this result implies that both branes end up interacting only weakly.
In Sec. \ref{S4}, the first post-Newtonian level parameters are
introduced and computed for the model. It will be found that they
depend on the shape of $\sigma(\phi)$ and on the positions of the
branes in the background. Then we explore the way in which
solar-system and binary-pulsar tests constrain the 5D configuration
for this class of models. Finally, in Sec. \ref{S5} we provide some
concluding remarks.

\section{BPS Braneworlds} \label{S2}

In this section we introduce BPS-brane models in some detail and
provide their low energy effective theory.

\subsection{The model}

\begin{figure}[ht]
\begin{center}
\includegraphics[width=0.45\textwidth]{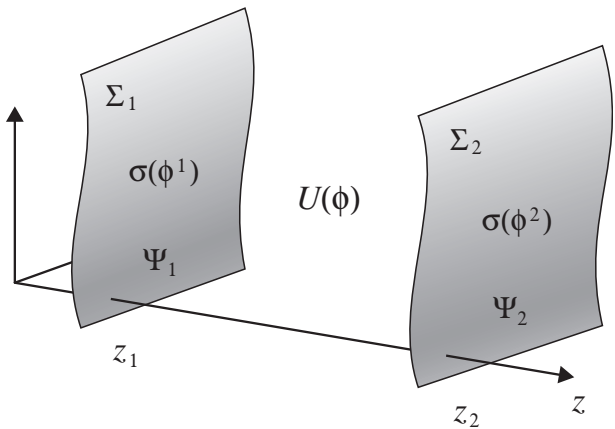}
\caption[Basic configuration]{Schematic representation of the
5-dimensional brane configuration. In the bulk there is a scalar
field $\phi$ with a bulk potential $U(\phi)$. Additionally, the
bulk-space is bounded by branes, $\Sigma_{1}$ and $\Sigma_{2}$,
with tensions $\lambda_{1} = + \sigma_{1}(\phi^{1})$ and
$\lambda_{2} = - \sigma_{2}(\phi^{2})$ respectively.} \label{F1}
\end{center}
\end{figure}

Let us consider a 5D manifold $M$ with topology $M = \mathbb{R}^{4}
\times S^{1}/ \mathbb{Z}_{2}$, where $\mathbb{R}^{4}$ is a fixed 4D
Lorentzian manifold without boundaries and $S^{1}/\mathbb{Z}_{2}$ is
the orbifold constructed from the one-dimensional circle with points
identified through a $Z_{2}$-symmetry. $M$ is bounded by two branes
located at the fixed points of $S^{1}/\mathbb{Z}_{2}$. Let us denote
the brane surfaces by $\Sigma_{1}$ and $\Sigma_{2}$ respectively and
the space $M$ bounded by the branes as the bulk space. In this model
there is a bulk scalar field $\phi$ with a bulk potential $U(\phi)$
and boundary values $\phi^{1}$ and $\phi^{2}$ at the branes. The
brane tensions are $\lambda_{1} = + \sigma(\phi^{1})$ and
$\lambda_{2} = - \sigma(\phi^{2})$ where $\sigma(\phi)$ is the brane
potential (sometimes called the superpotential). Additionally, we
consider the existence of matter fields $\Psi_{1}$ and $\Psi_{2}$
localized at the branes (see Fig. \ref{F1}).

The total action of the system is given by
\begin{eqnarray}
S = S_{\mathrm{bulk}} + S_{\mathrm{branes}} + S_{\mathrm{matter}},
\label{S2: tot-act}
\end{eqnarray}
where $S_{\mathrm{bulk}}$ is the action for the bulk fields,
including the gravitational field and the bulk scalar field
\begin{eqnarray}
S_{\mathrm{bulk}} = \frac{1}{2 \kappa_{5}^{2}} \int \!\!
\sqrt{-g^{(5)}} \, d^{\, 5}x \left[ R^{(5)} - \frac{3}{4} \big\{
(\partial \phi)^{2} + U \big\} \right].
\end{eqnarray}
Here, $g^{(5)}$ is the determinant of the five dimensional metric of
signature $(-,++++)$ and $R^{(5)}$ its Ricci scalar. The term
$S_{\Sigma}$ of Eq. (\ref{S2: tot-act}) is the action for the
boundary fields given by
\begin{eqnarray}
S_{\mathrm{branes}} = S_{\mathrm{GH}} + S_{\phi}^{\,1} +
S_{\phi}^{\,2},
\end{eqnarray}
where $S_{\mathrm{GH}}$ is the Gibbons-Hawking boundary term, and
$S_{\phi}^{\,1}$ and $S_{\phi}^{\,2}$ are the brane tensions terms
of the form
\begin{eqnarray} \label{eq: brane-tensions}
S_{\phi}^{\,1} &=& - \frac{3}{2 \kappa_{5}^{2}} \,
\int_{\Sigma_{1}}
\!\!\! d^{4} x \sqrt{-g} \,\, \sigma (\phi^{1}) , \\
S_{\phi}^{\,2} &=& + \frac{3}{2 \kappa_{5}^{2}} \,
\int_{\Sigma_{2}} \!\!\! d^{4} x \sqrt{-g} \,\, \sigma (\phi^{2}).
\end{eqnarray}
Finally, the last term in Eq. (\ref{S2: tot-act}) corresponds to the
action for the matter fields at the branes. It is given by:
\begin{eqnarray}
S_{\mathrm{matter}} = S_{1}[\Psi_{1},g_{\mu \nu}^{(1)}] +
S_{2}[\Psi_{2},g_{\mu \nu}^{(2)}],
\end{eqnarray}
where $\Psi_{1}$ and $\Psi_{2}$ denote the matter fields at each
brane, and $g_{\mu \nu}^{(1)}$ and $g_{\mu \nu}^{(2)}$ are the
respective 4D induced metrics.

To conclude, let us recall the BPS condition between the bulk
potential $U$ and the superpotential $\sigma$:
\begin{eqnarray} \label{S2: BPS-cond}
U= \left( \partial_{\phi} \sigma  \right)^{2} - \sigma^{2}.
\end{eqnarray}
Relation (\ref{S2: BPS-cond}) is of the utmost importance; it allows
the construction of static vacuum solutions in which the branes can
be located anywhere in the background (BPS states). When $\sigma$ is
a constant the Randall-Sundrum model is recovered with a negative
bulk cosmological constant $\Lambda_{5} = (3/8) U =  - (3/8) \,
\sigma^{2}$.

Suppose that $\rho$ is the matter energy density of the braneworld
universe, then, the low energy regime for this type of system is
characterized by the condition $\rho \ll \sigma \kappa_{5}^{2}$. In
the rest of this paper we assume that this is the case.

\subsection{5D geometry}

We now study the 5D geometry of the present setup. As already
mentioned, in the absence of matter, condition (\ref{S2: BPS-cond})
allows the construction of static vacuum solutions. This can be
outlined as follows: let us assume without loss of generality that
the 5D infinitesimal-metric $ds^{2}$ is given by
\begin{eqnarray}
ds^{2} = N^{2} dz^{2} + g_{\mu \nu} d x^{\mu} d x^{\nu}.
\end{eqnarray}
Here, the extra-dimension is parameterized by the coordinate $z$ and
$N$ is the scalar component of the 5D metric from the
four-dimensional point of view (it will be related to one of the
moduli on the branes). $g_{\mu \nu}$, on the other hand, is the
induced 4D metric at each $z$-slice parallel to the branes (the
branes are located at orbifold fixed points $z_{1}$ and $z_{2}$).
Now, consider the following factorization of $g_{\mu \nu}$:
\begin{eqnarray}
g_{\mu \nu} (z,x) = \omega^{2}(z) \, \tilde g_{\mu \nu}(x),
\label{S2: facto}
\end{eqnarray}
where $\tilde g_{\mu \nu}$ is a 4D metric satisfying the vacuum
Einstein's equation $\tilde G_{\mu \nu} = 0$ (here $\tilde G_{\mu
\nu}$ is the Einstein tensor constructed out of $\tilde g_{\mu
\nu}$). Then, condition (\ref{S2: BPS-cond}) allows us to find a
static vacuum solution to the entire system provided that the
following differential equations are fulfilled:
\begin{eqnarray}
\frac{\partial \phi}{\partial z} &=& N \frac{\partial
\sigma}{\partial \phi}, \label{S2: match1} \\
\frac{1}{\omega} \frac{\partial \omega}{\partial z} &=& -
\frac{1}{4} N \sigma. \label{S2: match2}
\end{eqnarray}
In this solution the branes can be localized anywhere in the 5D
background without modifying the geometry around them. In what
follows we deduce a few important results out of relations (\ref{S2:
match1}) and (\ref{S2: match2}).

\subsubsection{Parameterization of the extra-dimension}

An immediate consequence of Eq. (\ref{S2: match1}) is the following:
if $\sigma$ is not a constant function of $\phi$ then we can
parameterize the coordinate $z$ in terms of $\phi$-values:
\begin{eqnarray}
N dz = \left(
\partial_{\phi} \sigma \right)^{-1} \! d \phi. \label{S2: dz-dpsi}
\end{eqnarray}
Note that this parameterization depends heavily on the shape of
$\sigma(\phi)$. In particular, observe that as $\sigma (\phi)$
approaches a local maximum or minimum, the $z$-coordinate tends to
infinity. This implies that $\partial_{\phi} \sigma$ cannot change
signs in the entire bulk space. This observation simplifies much of
the analysis on the braneworld geometry. Without lost of generality
we may assume $\partial_{\phi} \sigma \geqslant 0$ everywhere in
$M$, thus $\phi$ becomes an increasing monotonic function of $z$
(otherwise we can perform the change of variables $\phi \rightarrow
- \phi$).

\subsubsection{Profile of the warp factor $\omega$}

\begin{figure}[ht]
\begin{center}
\includegraphics[width=0.45\textwidth]{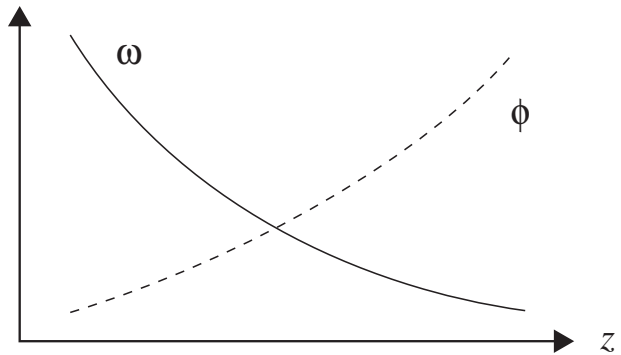}
\caption[Behavior of the fields]{The figure shows the generic
behavior of the warp factor $\omega$ and the scalar field $\phi$
as functions of $z$. We can always choose $\phi$ to increase and
$\omega$ to decrease in the $z$-direction.} \label{F2}
\end{center}
\end{figure}

Observe from Eq. (\ref{S2: match2}) that the 5D geometry profile is
dictated by the shape of $\sigma$; if $\sigma > 0$ ($\sigma < 0$)
then the warp factor $\omega$ is a decreasing (increasing) function
of $z$. In Sec. \ref{S3} we shall see that in order to have
consistent dynamics at the branes, $\sigma$ must have the same sign
everywhere in the bulk. In general, this implies that the brane
tensions $\lambda_{1}$ and $\lambda_{2}$ must have opposite signs
(as in the Randall-Sundrum model). Again, without lost of
generality, we may assume that $\sigma > 0$ everywhere in the bulk
(otherwise, we can invert the positions of the branes). Fig.
\ref{F2} shows the generic behavior for $\omega$ and $\phi$ as
functions of $z$.

Putting Eqs. (\ref{S2: match1}) and (\ref{S2: match2}) together,
$\omega(z)$ can be solved and expressed in terms of $\phi(z)$ in an
$N$-independent way:
\begin{eqnarray}
\omega(z) &=& \exp \left[ -\frac{1}{4} \int_{\phi_{*}}^{\phi(z)}
\!\! \alpha^{-1}(\phi) \, d \phi \right], \label{S2: omega} \\
\alpha(\phi) &=& \frac{1}{\sigma} \frac{\partial \sigma}{\partial
\phi}. \label{alpha def}
\end{eqnarray}
Here $\phi_{*}$ is an arbitrary constant value for $\phi$. If we
define $\omega_{1} = \omega(z_{1})$ and $\omega_{2} =
\omega(z_{2})$, then the induced metrics to the first and second
branes are $\omega^{2}_{1} \, \tilde g_{\mu \nu}$ and
$\omega^{2}_{2} \, \tilde g_{\mu \nu}$ respectively. Furthermore,
both metrics are conformally related by the warp factor
\begin{eqnarray}
\omega_{2} / \omega_{1} = \exp \left[ -\frac{1}{4}
\int_{\phi^{1}}^{\phi^{2}} \!\! \alpha^{-1} \, d \phi \right].
\label{S2: warp}
\end{eqnarray}
The parameter $\alpha$ defined in Eq. (\ref{alpha def}) will have an
important role in the rest of the paper.

\subsubsection{Existence of singularities} \label{Sing}

Finally, consider the possibility of having singularities at a
finite position in the bulk. Singularities appear in the bulk either
if $\omega(z_{1}) = + \infty$ or $\omega(z_{2})=0$. Thus, Eq.
(\ref{S2: warp}) tells us that singularities exist provided that the
following integral diverges:
\begin{eqnarray} \label{S2: integr-div}
\int_{\phi^{1}}^{\phi^{2}} \!\!\!\!\! \alpha^{-1}(\phi) \, d \phi
\rightarrow + \infty.
\end{eqnarray}
Recall that $\sigma$ and $\alpha$ are positive in the entire domain
$(\phi^{1} , \phi^{2})$ meaning that singularities can only appear
if $\alpha^{-1} = \sigma \, (\partial_{\phi} \sigma)^{-1}$ diverges
at any of the two boundary values $\phi^{1}$ or $\phi^{2}$. Since
the case $\partial_{\phi} \sigma = 0$ corresponds to $z = \pm
\infty$, we conclude that the only possible bulk-singularities are
those where $\sigma(\phi^{2}) = + \infty$. This corresponds to
$\omega_{2} = 0$. From Eq. (\ref{S2: dz-dpsi}) we see that for such
a singularity to be located in the bulk at a finite distance from
the first brane $\Sigma_{1}$, the following additional condition
must hold:
\begin{eqnarray} \label{S2: cond-sing}
0 < \int_{\phi^{1}}^{\phi^{2}} \left( \frac{\partial
\sigma}{\partial \phi} \right)^{\!\! -1} d \phi < + \infty.
\end{eqnarray}
It should be mentioned that in this analysis $\phi^{2}$ is allowed
to reach arbitrary large values.

\subsection{4D effective theory}

We finish this section with the introduction of the low energy
effective action describing the system (see \cite{Palma-Davis1,
Palma-Davis2} for a detailed study on BPS-braneworld effective
theories). For the present discussion it will be convenient to
display the theory in two different (but equivalent) frames: The
``background'' frame, where the gravitational field $\tilde g_{\mu
\nu}$ is the same one of Eq. (\ref{S2: facto}); and the Einstein's
frame, where the term (in the action) containing the Ricci scalar is
independent of the moduli $\phi^{1}$ and $\phi^{2}$. Before
continuing, please notice that in the following discussion we shall
use the warp factors $\omega_{1}$ and $\omega_{2}$ instead of the
boundaries $\phi^{1}$ and $\phi^{2}$ to parameterize the
moduli-space [the relation between these two pairs is easily
obtained from Eq. (\ref{S2: omega})].

\subsubsection{Background frame}

In this frame, the low energy effective action for the
$\omega$-fields reads:
\begin{eqnarray}
S &=& \frac{3}{4 \pi G_{*}} \int d^{4}x \sqrt{- \tilde g}
\bigg[ \frac{W^{2}}{12} \tilde R + \frac{k}{\sigma_{1}} (\tilde
\partial \omega_{1})^{2} -
\frac{k}{\sigma_{2}} (\tilde \partial \omega_{2})^{2} \bigg]  \nonumber\\
&& + S_{1}[\Psi_{1}, \omega^{2}_{1} \, \tilde g_{\mu \nu}] +
S_{2}[\Psi_{2}, \omega^{2}_{2} \, \tilde g_{\mu \nu}], \qquad
\label{S1: EFF Action1}
\end{eqnarray}
where the factor $W^{2} > 0$ is given by
\begin{eqnarray}
W^{2} = - 4 k \int_{\omega_{1}}^{\omega_{2}} \frac{\omega}{\sigma}
\,\, d \omega . \label{S2: W2}
\end{eqnarray}
In the previous expressions, the coefficient $k$ is an arbitrary
positive constant with dimensions of mass included to make $W^{2}$
dimensionless. The action contains now the dimensionful constant
$G_{*} = k \kappa_{5}^{2}/16 \pi$, which should not be identified
with the Newton's constant $G_{N}$ (we will clarify the relation
between $G_{*}$ and $G_{N}$ in Sec. \ref{edd def}). We have also
defined $\sigma_{1} = \sigma(\phi^{1})$ and $\sigma_{2} =
\sigma(\phi^{2})$ for simplicity. Observe that the kinetic term for
$\omega_{1}$ appears to have the wrong sign; this is an artifact of
the background frame and has no physical significance for stability
analysis.

\subsubsection{Einstein's frame}

To obtain the Einstein's frame action it is necessary to perform the
conformal transformation $\tilde g_{\mu \nu} \rightarrow g_{\mu \nu}
= W^{2} \, \tilde g_{\mu \nu}$ in Eq. (\ref{S1: EFF Action1}). We
find:
\begin{eqnarray}
S &=& \frac{1}{4 \pi G_{*}} \int d^{4}x \sqrt{- g} \bigg[
\frac{1}{4} R - \frac{1}{2} g^{\mu \nu} \gamma^{a b}
\partial_{\mu} \omega_{a} \partial_{\nu} \omega_{b} \bigg]  \nonumber\\
&& + S_{1}[\Psi_{1}, A_{1}^{2}  g_{\mu \nu}] + S_{2}[\Psi_{2},
A_{2}^{2}  g_{\mu \nu}], \qquad \label{S2: EFF Action2}
\end{eqnarray}
where the matter-gravity couplings $A_{1}^{2}$ and $A_{2}^{2}$ are
given by
\begin{eqnarray}
A_{1}^{2} = \frac{\omega_{1}^{2}}{W^{2}}, \qquad A_{2}^{2} =
\frac{\omega_{2}^{2}}{W^{2}},
\end{eqnarray}
and the symmetric sigma-model metric $\gamma^{a b}$ by
\begin{eqnarray}
\gamma^{1 1} &=& - \frac{6 k }{W^{2} \sigma_{1}}
\bigg[ 1 - \frac{2k A_{1}^{2}}{\sigma_{1}} \bigg] ,  \\
\gamma^{2 2} &=& + \frac{6 k }{W^{2} \sigma_{2}}
\bigg[ 1 + \frac{2k A_{2}^{2}}{\sigma_{2}} \bigg] , \\
\gamma^{1 2} &=& \gamma^{2 1} = - \frac{12 k^{2} \omega_{1}
\omega_{2}}{W^{4} \sigma_{1} \sigma_{2}}.
\end{eqnarray}
It is possible to show that $\gamma^{a b}$ is positive-definite
($\gamma^{11}$ is positive) and that the inverse metric $\gamma_{a b}$
has components:
\begin{eqnarray}
\gamma_{1 1} &=& + \frac{W^{4} \sigma_{1}}{6 k K^{2}}
\bigg[ 1 + \frac{2k A_{2}^{2}}{\sigma_{2}} \bigg] ,  \\
\gamma_{2 2} &=& - \frac{W^{4} \sigma_{2}}{6 k K^{2}}
 \bigg[ 1 - \frac{2k A_{1}^{2}}{\sigma_{1}} \bigg], \\
\gamma_{1 2} &=& \gamma_{2 1} = + \frac{ W^{2} \omega_{1}
\omega_{2}}{3 K^{2}},
\end{eqnarray}
where $K^{2} > 0$ has been defined as
\begin{eqnarray}
K^{2} = -8k \int_{\omega_{1}}^{\omega_{2}} \alpha^{2}
\frac{\omega}{\sigma} \, d \omega. \label{S2: K2}
\end{eqnarray}
It is a simple matter to show that $K^{2}$ and $W^{2}$ are related
by
\begin{eqnarray}
K^{2} + W^{2} =  2k \frac{\omega_{1}^{2}}{\sigma_{1}} - 2k \frac{
\omega_{2}^{2}}{\sigma_{2}}  .
\end{eqnarray}
Observe that the positivity of $W^{2}$ and $K^{2}$ is independent of
any conventions on the signs of $\sigma$ or $\partial_{\phi} \sigma$
[to see this, it is enough to use Eq. (\ref{S2: match1}) in the
definitions of both quantities]. In Appendix \ref{A1} we provide the
connections $\gamma^{a b}_{c}$ of the present sigma-model.

\subsubsection{Including departures from the BPS condition}

Although we are not concerned with the presence of a dark energy
term in the 4D effective theory, here we consider departures from
the BPS condition (\ref{S2: BPS-cond}) for completeness. If we
assume, without loss of generality, a shift in the tensions
$\lambda_{1} \rightarrow \lambda_{1} = \sigma_{1} + v_{1}$ and
$\lambda_{2} \rightarrow
 \lambda_{2} = - \sigma_{2} + v_{2}$, then the following potential emerges in
Eq. (\ref{S2: EFF Action2}):
\begin{eqnarray}
V = \frac{3k}{4} \left[ A_{1}^{4} v_{1} -  A_{2}^{4} v_{2}
\right].
\end{eqnarray}
(A shift of the bulk potential can be absorbed by a redefinition
of $\sigma$).

\section{Braneworld dynamics} \label{S3}

In this section we analyze the equations of motion for the system
and study the cosmological evolution of the branes. The simplest
frame to deduce and analyze these equations is the background-frame.
Here, the evolution of the warp factors $\omega_{1}$ and
$\omega_{2}$ can be interpreted without difficulty as the movement
of branes on the 5D background.

\subsection{Matter energy-momentum tensor}

Before deducing the equations of motion it will be helpful to define
the 4D energy-momentum tensor $T^{a}_{\mu \nu}$ for the matter
content at the brane $\Sigma_{a}$ in the most appropriate way. We
proceed as follows:
\begin{eqnarray}
T_{\mu \nu}^{a} = - 2 \frac{1}{\sqrt{- g_{a}}}\frac{\delta
S_{a}}{\delta g^{\mu \nu}_{a}}, \label{S3: T}
\end{eqnarray}
where $g_{\mu \nu}^{a} = \omega^{2}_{a} \, \tilde g_{\mu \nu}$.
Accordingly, it is also convenient to define the trace $T_{a}$
using $g_{\mu \nu}^{a}$:
\begin{eqnarray}
T_{a} =  g^{\mu \nu}_{a} T_{\mu \nu}^{a} = \omega^{-2}_{a} \,
\tilde g^{\mu \nu} T_{\mu \nu}^{a}. \label{S3: Trace}
\end{eqnarray}
These definitions are consistent with conservation of the
energy-momentum tensor at each brane in the following sense:
\begin{eqnarray}
\nabla_{a}^{\mu} T^{a}_{\mu \nu} = 0, \label{S3: cons}
\end{eqnarray}
where $\nabla_{a}^{\mu}$ is the covariant derivative constructed out
of $g_{\mu \nu}^{a}$. Nevertheless, we should keep in mind that the
emission of gravity waves may induce a nonconservation term in
(\ref{S3: cons}).

\subsection{Equations of motion}

Varying the action (\ref{S1: EFF Action1}) in terms of the moduli,
we obtain the following equations of motion:
\begin{eqnarray}
\tilde \Box \omega_{1} = - 2 \frac{\alpha_{1}^{2}}{\omega_{1}}
(\tilde
\partial \omega_{1})^{2} + \frac{1}{6} \omega_{1} \tilde R +
\frac{\kappa_{5}^{2}}{24} \omega_{1}^{3} \sigma_{1} \, T_{1},
\end{eqnarray}
and
\begin{eqnarray}
\tilde \Box \omega_{2} = - 2 \frac{\alpha_{2}^{2}}{\omega_{2}}
(\tilde
\partial \omega_{2})^{2} + \frac{1}{6} \omega_{2} \tilde R -
\frac{\kappa_{5}^{2}}{24} \omega_{2}^{3} \sigma_{2} \, T_{2}.
\end{eqnarray}
On the other hand, varying (\ref{S1: EFF Action1}) with respect to
$\tilde g_{\mu \nu}$, we find:
\begin{eqnarray}
W^{2} \tilde R_{\mu \nu} &=& \tilde \nabla_{\mu} \tilde
\nabla_{\nu}
W^{2} + \frac{1}{2} \tilde g_{\mu \nu} \tilde \Box W^{2} \nonumber\\
&& + 12 k \left[ \frac{1}{\sigma_{1}}(\tilde \partial
\omega_{1})^{2}
- \frac{1}{\sigma_{2}}(\tilde \partial \omega_{2})^{2}    \right]  \nonumber\\
&& + 8 \pi G_{*} \sum_{a} \omega_{a}^{2} \left[
T_{\mu \nu}^{a} - \frac{1}{2} \, \omega_{a}^{2} \tilde g_{\mu \nu}
T_{a} \right].
\end{eqnarray}
Recall that $16 \pi G_{*} = k \kappa_{5}^{2}$.
The previous set of equations can be combined together to obtain
the following simple but powerful equation:
\begin{eqnarray}
K^{2} \tilde R = 24 k \left[
\frac{\alpha_{2}^{2}}{\sigma_{2}}(\tilde
\partial \omega_{2})^{2} - \frac{\alpha_{1}^{2}}{\sigma_{1}}(\tilde \partial
\omega_{1})^{2}  \right]. \label{S3: K2R}
\end{eqnarray}
Here it is possible to see that $\sigma$ must always have the same
sign within $(\phi^{1}, \phi^{2})$. Otherwise $K^{2}$ would diverge
and Eq. (\ref{S3: K2R}) would be senseless. [To see that $K^{2}$
diverges it is enough to express the integral (\ref{S2: K2}) in
terms of $\phi$ and study its linear behavior around $\sigma = 0$].

\subsection{Cosmological equations}

Let us now consider the cosmological evolution of the branes. Assume
that the effective fields are functions of time only and that the 4D
metric has a flat Friedmann-Robinson-Walker profile
\begin{eqnarray}
ds^{2} = a^{2}(t) \left( -dt^{2} + dr^{2} \right),
\end{eqnarray}
where $a(t)$ is the scale factor and $t$ the conformal time in the
background frame. The nonvanishing components of the matter
energy-momentum tensor (\ref{S3: T}) can be expressed in terms of
the energy density $\rho_{a}$ and pressure $p_{a}$ (as measured by
local observers at the branes)
\begin{eqnarray}
T_{00}^{b} = (a \omega_{b})^{2} \rho_{b}, \quad \mathrm{and} \quad
T_{ij}^{b} = \delta_{ij} (a \omega_{b})^{2} p_{\, b},
\end{eqnarray}
and, by defining the equation of state $p_{a} = s_{a} \rho_{a}$, we
find the trace (\ref{S3: Trace}) to have the form
\begin{eqnarray}
T_{a} = - (1 - 3 s_{a}) \rho_{a}.
\end{eqnarray}
Additionally, from the energy-momentum conservation relation
(\ref{S3: cons}) it is possible to show that $\rho_{b} \propto (a
\omega_{b})^{- 3 (1 + s_{b})}$. Then, the scalar fields
are found to satisfy the following equations of motion:
\begin{eqnarray}
\ddot \omega_{1} + 2 H \dot \omega_{1}  &=& - 2
\frac{\alpha_{1}^{2}}{\omega_{1}} (\dot \omega_{1})^{2} -
\frac{1}{6} \omega_{1} a^{2} \tilde R \nonumber\\ && +
\frac{\kappa_{5}^{2}}{24} \omega_{1}^{3} \sigma_{1} a^{2}  (1-3 s_{1})
\rho_{1}, \label{S3: omega1}
\end{eqnarray}
and
\begin{eqnarray}
\ddot \omega_{2} + 2 H \dot \omega_{2}  &=& - 2
\frac{\alpha_{2}^{2}}{\omega_{2}} (\dot \omega_{2})^{2} -
\frac{1}{6} \omega_{2} a^{2} \tilde R \nonumber\\ && -
\frac{\kappa_{5}^{2}}{24} \omega_{2}^{3} \sigma_{2} a^{2}  (1-3 s_{2})
\rho_{2}, \label{S3: omega2}
\end{eqnarray}
where $H = \dot a/a$ and $\tilde R = 6 \, \ddot a / a^{3}$. On the
other hand, the Friedmann's equation is found to be
\begin{eqnarray}
2k \frac{\omega_{1}^{2}}{\sigma_{1}} \left( H + \frac{\dot
\omega_{1}}{\omega_{1}} \right)^{2} - 2k
\frac{\omega_{2}^{2}}{\sigma_{2}} \left( H + \frac{\dot
\omega_{2}}{\omega_{2}} \right)^{2} \nonumber\\ = K^{2} H^{2}
+\frac{8 \pi G_{*}}{3} \, a^{2} \left( \omega_{1}^{4} \rho_{1} +
\omega_{2}^{4} \rho_{2} \right). \label{S3: Friedmann}
\end{eqnarray}
Finally, it is useful to recast Eq. (\ref{S3: K2R}) for the
cosmological case
\begin{eqnarray}
a^{2} K^{2} \tilde R = 24 k \left[
\frac{\alpha_{1}^{2}}{\sigma_{1}}(\dot \omega_{1})^{2} -
\frac{\alpha_{2}^{2}}{\sigma_{2}}( \dot \omega_{2})^{2} \right].
\label{S3: K2R-cos}
\end{eqnarray}

\subsection{Cosmological evolution}

\begin{figure}[ht]
\begin{center}
\includegraphics[width=0.45\textwidth]{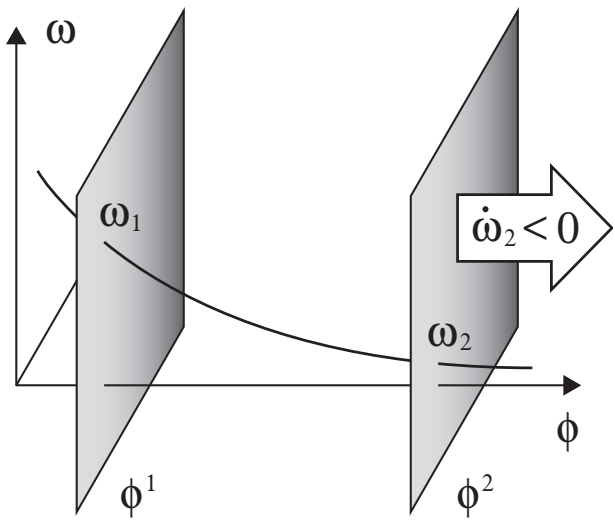}
\caption[Behavior of the fields]{The figure shows the evolution of
the second brane for the initial condition $\dot \omega_{2} < 0$.
The second brane is forced to evolve towards the fixed point
$(\omega_{2}, \dot \omega_{2}) = 0$ regardless of the evolution of
the first brane.} \label{F3}
\end{center}
\end{figure}

We now analyze the evolution of cosmic branes on the background.
Finding solutions to the system of Eqs. (\ref{S3: omega1}),
(\ref{S3: omega2}) and (\ref{S3: Friedmann}) is in general difficult
without the use of numerical methods. However, it is possible to
show that the system has a generic attractor solution where both
branes decouple; let us assume an arbitrary configuration in which
$\dot \omega_{2} < 0$. This corresponds to have the second brane
$\Sigma_{2}$ moving to the right (see Fig. \ref{F3}).
Then, regardless the evolution of the first brane, the second brane
will always continue moving to the right ($\dot \omega_{2} < 0$ for
all times). This can be seen most easily by rewriting Eq. (\ref{S3:
omega2}) in the form
\begin{eqnarray}
 2 H \dot \omega_{2}  &=& - \ddot \omega_{2} - 2
\frac{\alpha_{2}^{2}}{\omega_{2}} (\dot \omega_{2})^{2} -
\frac{1}{6} \omega_{2} a^{2} \tilde R \nonumber\\ && -
\frac{\kappa_{5}^{2}}{24} \omega_{2}^{3} \sigma_{2} a^{2}  (1-3 s_{2})
\rho_{2}, \label{S3: omega2-n1}
\end{eqnarray}
and proving that the right hand side remains always negative:
Observe that for $\dot \omega_{2}$ to change signs it is necessary
to have $\ddot \omega_{2}$ positive. Note also, from Eq. (\ref{S3:
K2R-cos}), that near $\dot \omega_{2} = 0$ the Ricci scalar $\tilde
R$ becomes positive. From these two observations we deduce that the
right hand side of (\ref{S3: omega2-n1}) is always negative (here we
are making the additional assumption that $s_{2} \leqslant 1/3$,
which includes matter, radiation, and cosmological constant
dominated universes). The later is an important result for BPS
braneworlds: in the low energy regime, the second brane is generally
expected to evolve towards the fixed point $(\omega_{2}, \dot
\omega_{2}) = 0$.

\subsection{Decoupled branes}

Notice that if the second brane has effectively evolved to a
configuration near the fixed point $(\omega_{2}, \dot \omega_{2}) =
0$, then the first brane evolution becomes independent of the second
brane $\Sigma_{2}$. As a matter of fact, the Friedmann's Eq.
(\ref{S3: Friedmann}) reduces to
\begin{eqnarray}
 \left( H + \frac{\dot
\omega_{1}}{\omega_{1}} \right)^{2} - \frac{\sigma_{1} K^{2}
H^{2}}{2k \omega_{1}^{2}} =  a^{2} \omega_{1}^{2}
\frac{\kappa_{5}^{2}}{12} \, \sigma_{1} \rho_{1} . \label{S3:
Friedmann2}
\end{eqnarray}
(In the next section we shall see that current observations can further
constrain the form of this equation). More generally, the first brane dynamics is
affected only by the moduli $\omega_{1}$, and the braneworld system
becomes described by the following single scalar-tensor theory:
\begin{eqnarray} \label{onebrane}
S &=& \frac{3}{4 \pi G_{*}} \int d^{4}x \sqrt{- \tilde g}
\bigg[ \frac{W^{2}}{12} \tilde R + \frac{k}{\sigma_{1}} (\tilde
\partial \omega_{1})^{2} \bigg]  \nonumber\\
&& + S_{1}[\Psi_{1}, \omega^{2}_{1} \, \tilde g_{\mu \nu}],
\end{eqnarray}
where now the factor $W^{2}$ is given by:
\begin{eqnarray}
W^{2} = 4 k \int_{0}^{\omega_{1}} \frac{\omega}{\sigma}
\,\, d \omega .
\end{eqnarray}
Let us emphasize here that $\omega_{2}=0$ could be either a
singularity (as discussed in Sec. \ref{Sing}) or an asymptotic state
of the system.

\section{Observational constraints} \label{S4}

General Relativity is characterized for having constant couplings
$A_{1}$ and $A_{2}$ [see Eq. (\ref{S2: EFF Action2})]. The
parameterized post-Newtonian formalism for bi-scalar-tensor theories
consists of analyzing the effects of varying couplings $A_{1}$ and
$A_{2}$ on the gravitational interaction between different bodies.
This analysis is usually done by studying relativistic corrections
in powers of $1/c^{2}$ to Newton's gravitational potential.

In this article we consider the first post-Newtonian (1PN) level,
where the analysis is considered up to $1/c^{2}$ order. Also, we
shall be concerned with interacting bodies localized at the same
brane (let us say the first brane $\Sigma_{1}$), thus we only
consider the effects of the coupling $A_{1}$. Recall that in Sec.
\ref{S3} we saw that the second brane $\Sigma_{2}$ is generally
expected to evolve towards the fixed point ($\omega_{2}, \dot
\omega_{2}) = 0$. In particular, this is the case for branes
dominated by matter and radiation, so we may safely assume that at
late cosmological times both branes are far enough apart and do not
interact through scalar interchange (this has been verified
numerically for the particular case of dilatonic braneworlds
\cite{num-dil}).

\subsection{Eddington parameters} \label{edd def}

Let us start our analysis with a brief outline on the definition of
the Eddington parameters. Consider the interaction between bodies
localized at the first brane (and therefore affected by the coupling
$A_{1}$) and express $\omega_{1}$ and $\omega_{2}$ about their
cosmological values $\omega^{0}_{1}$ and $\omega^{0}_{2}$ as
\begin{eqnarray}
\omega_{1} = \omega^{0}_{1} + \varphi_{1}, \qquad \mathrm{and}
\qquad \omega_{2} = \omega^{0}_{2} + \varphi_{2}.
\end{eqnarray}
Then, $A_{1}$ can be expanded in the form
\begin{eqnarray}
A_{1} = A_{1}^{0} \left[ 1 + \eta^{a}_{0} \varphi_{a} +
\frac{1}{2} (\eta^{a}_{0} \eta^{b}_{0} + \beta^{a b}_{0})
\varphi_{a} \varphi_{b} + \ldots \right],
\end{eqnarray}
where $\eta^{a}_{0}$ and $\beta^{a b}_{0}$ are the background
cosmological values of the following quantities:
\begin{eqnarray}
\eta^{a} = \frac{1}{A_{1}} \frac{\partial A_{1}}{\partial
\omega_{a}} , \qquad \mathrm{and} \qquad \beta^{a b} = D^{a}
\eta^{b},
\end{eqnarray}
where $D^{a} \eta^{b} = \partial^{a} \eta^{b} - \gamma^{a b}_{c}
\eta^{c}$ is the sigma-model covariant derivative of $\eta^{a}$ with
connections $\gamma^{a b}_{c}$ (see Appendix \ref{A1}). From here on
we omit the ``$0$'' label denoting quantities evaluated at their
cosmological background values. The effects of $A_{1}$ on the
gravitational interaction can now be studied in terms of
$\varphi$-scalars exchange between bodies. For instance, if we
compute the Schwarzchild metric for a body of mass $m$ taking into
account the perturbations $\varphi_{1}$ and $\varphi_{2}$, we find
\begin{eqnarray}
- g_{0 0} &=& 1 - 2 \frac{G_{N} m}{r} + 2 \beta \left( \frac{G_{N}
m}{r}\right)^{2} + \mathcal{O} \left( 1/c^{6} \right), \quad \\
g_{i j} &=& \delta_{i j} \left[ 1 + 2 \gamma \frac{G_{N} m}{r} \right]
+ \mathcal{O} \left( 1/c^{4} \right),
\end{eqnarray}
where $G_{N} = G_{*} A_{1}^{2}$ is the Newton's constant as
measured by Cavendish experiments in the brane.
In the previous equations we have introduced
the two Eddington parameters $\beta$ and
$\gamma$ as \cite{PN1}:
\begin{eqnarray}
\gamma - 1 &=& - \frac{2 \eta^{2}}{1 + \eta^{2}}, \label{S4:
gamma}\\
\beta - 1 &=& \frac{1}{2} \frac{\beta^{a b} \eta_{a} \eta_{b}}{(1
+ \eta^{2})^{2}}, \label{S4: beta}
\end{eqnarray}
where $\eta^{2} = \eta_{a} \eta^{a}$ (indexes are raised and
lowered by the sigma model metrics $\gamma^{a b}$ and $\gamma_{a
b}$). The basic role of these quantities is to parameterize
deviations to Einstein gravity at the first post-Newtonian level
(see Appendix \ref{2PN} for the definition of the second
post-Newtonian level parameters).

\subsection{Braneworld's Eddington parameters}

In the specific case of our model, we can compute $\eta^{2}$ and
$\beta^{a b} \eta_{a} \eta_{b}$ in terms of $\sigma$ and other
quantities
\begin{eqnarray}
\eta^{2} &=& \frac{1}{3} \left[ 1 - \frac{\sigma_{1}}{2 k
A_{1}^{2}}  \right] , \label{S4: eta}\\
\beta^{a b} \eta_{a} \eta_{b} &=& \frac{1}{9} \left(1 - 3 \eta^{2}
\right) \left[ 3 \eta^{2} - 2 \alpha_{1}^{2} \left(1 - 3 \eta^{2}
\right) \right] .
\end{eqnarray}
Before continuing the discussion, it will be useful to introduce a
new parameter encoding information from the bulk geometry. Recalling
the definitions of $W^{2}$ and $K^{2}$ in Eqs. (\ref{S2: W2}) and
(\ref{S2: K2}) respectively, we define
\begin{eqnarray}
\langle \alpha^{2} \rangle = \frac{1}{2} \frac{K^{2}}{W^{2}}.
\end{eqnarray}
Here $\langle \alpha^{2} \rangle$ can be interpreted as the average
value of $\alpha^{2}$ over the bulk taking into account the
background geometry; for instance, if the background's geometry is
strongly warped near $\Sigma_{1}$, we should expect $\langle
\alpha^{2} \rangle$  to be small and close to the boundary value
$\alpha_{1}^{2}$. In particular, if $\alpha$ is constant, then
$\langle \alpha^{2} \rangle = \alpha_{1}^{2} = \alpha^{2}$, which
corresponds to dilatonic braneworlds. We shall come back to the
geometrical meaning of $\langle \alpha^{2} \rangle$ in Sec.
\ref{geom}.

To further simplify our analysis, observe from Eq. (\ref{S4: gamma})
that if $\gamma \simeq 1$ then $\eta^{2} \simeq 0$. Thus, neglecting
second order terms in $\eta^{2}$ on Eqs. (\ref{S4: gamma}) and
(\ref{S4: beta}), we can reexpress the post-Newtonian parameters
simply as
\begin{eqnarray}
1 - \gamma &=& \frac{4}{3} \langle \alpha^{2} \rangle +
\frac{2}{3}
\frac{\omega^{2}_{2} \sigma_{1}}{\omega^{2}_{1} \sigma_{2}},
\label{S4: brane-edd1} \\
1 - \beta &=& \frac{1}{9} \left( \alpha_{1}^{2} -   \langle
\alpha^{2} \rangle  \right) - \frac{1}{18} \frac{\omega^{2}_{2}
\sigma_{1}}{\omega^{2}_{1} \sigma_{2}} \label{S4: brane-edd2}.
\end{eqnarray}
Notice that $\gamma$ is related to the global configuration of the
system, while $\beta$ encodes departures of the brane parameter
$\alpha_{1}^{2}$ from $\langle \alpha^{2} \rangle$. In particular,
if the branes are far apart, we find:
\begin{eqnarray}
1 - \gamma &=& \frac{4}{3} \langle \alpha^{2} \rangle,
\label{S4: brane-edd3}\\
1 - \beta &=& \frac{1}{9} \left( \alpha_{1}^{2} -   \langle
\alpha^{2} \rangle  \right) \label{S4: brane-edd4}.
\end{eqnarray}
In what follows, we study the current observational constraints on these
parameters.

\subsection{Solar-system tests}

Current solar-system tests place tight constraints on the 1PN level
parameters. A few of the most relevant tests constraining these
parameters are: The observations on the perihelion shift of Mercury,
which implies \cite{Mercury}
\begin{eqnarray}
| 2 \gamma - \beta - 1 | < 3 \times 10^{-3}, \label{S5: Mercury}
\end{eqnarray}
the lunar laser ranging experiment, which provides the value
\cite{Lunar}
\begin{eqnarray}
4 \beta - \gamma - 3  = (-0.7 \pm 1) \times 10^{-3}, \label{S5:
LLR}
\end{eqnarray}
the light deflection measured by the very long baseline
interferometry experiment \cite{VLB}
\begin{eqnarray}
|\gamma - 1 | < 4 \times 10^{-4}, \label{S5: Light}
\end{eqnarray}
and the measurement of the time delay variation to the Cassini
spacecraft near the solar conjunction, which gives the impressive
result \cite{Cassini}
\begin{eqnarray}
\gamma - 1  = (2.1 \pm 2.3) \times 10^{-5}. \label{S5: Cassini}
\end{eqnarray}
In the present analysis it will be enough to consider only the
constraint shown in  Eq. (\ref{S5: Cassini}). Furthermore, since in
our theory $1 - \gamma$ is always positive, we can reinterpret this
bound as $1 - \gamma  < 2 \times 10^{-6}$. This imposes the
following two inequalities
\begin{eqnarray}
\frac{\omega^{2}_{2} \sigma_{1}}{\omega^{2}_{1} \sigma_{2}} &<& 3
\times 10^{-6} , \label{S5: omega-1-2} \\
\langle \alpha^{2} \rangle &<& 1.5 \times 10^{-6}. \label{S5:
alpha}
\end{eqnarray}
Bound (\ref{S5: omega-1-2}) gives us information regarding the
relative positions of both branes on the background, however it does
not constrain the geometry of the system. For example, if
$\sigma(\phi)$ is a slowly varying function of $\phi$ in the bulk
(between both branes), then the warp factor $\omega (z)$ must vary
quickly from one brane to the other (a strongly warped geometry).
Nevertheless, it may be possible to have configurations where the
warp factor does not necessarily vary much from one brane to the
other, but $\sigma(\phi)$ does. In any case, it is important to note
that Eq. (\ref{S5: omega-1-2}) is in good agreement with our
analysis made on the cosmological evolution of the branes, where we
found that the second brane has an attractor behavior to the state
$(\omega_{2}, \dot \omega_{2}) = 0$. In what follows we assume that
the terms involving $\omega_{2}$ in Eqs. (\ref{S4: brane-edd1}) and
(\ref{S4: brane-edd2}) can be safely neglected, and continue using
Eqs. (\ref{S4: brane-edd3}) and (\ref{S4: brane-edd4}) instead.

Let us now move to inequality (\ref{S5: alpha}). This bound
constrains the average value of $\alpha^{2}$ in the bulk, and can be
used to discard many possible shapes for the brane-potential
$\sigma(\phi)$. Furthermore, observe that the Friedmann's Eq.
(\ref{S3: Friedmann2}) can be rewritten in the form
\begin{eqnarray}
 \left( H + \frac{\dot
\omega_{1}}{\omega_{1}} \right)^{2} = 2 \, \langle \alpha^{2} \rangle
H^{2} + a^{2} \omega_{1}^{2} \frac{\kappa_{5}^{2}}{12} \, \sigma_{1}
\rho_{1} . \label{S3: Friedmann3}
\end{eqnarray}
Neglecting the term proportional to $\langle \alpha^{2} \rangle$ and
defining the proper Hubble parameter $\mathcal{H} \equiv (H + \dot
\omega_{1}/\omega_{1})/a^{2} \omega_{1}^{2}$ (as it would be
measured by an observer at the $\Sigma_{1}$ brane frame), we arrive
at the simple expression
\begin{eqnarray}
\mathcal{H}^{2} = \frac{\kappa_{5}^{2}}{12} \, \sigma_{1} \rho_{1}
= \frac{8 \pi}{3} G_{N} \rho_{1} , \label{S3: Friedmann4}
\end{eqnarray}
where we have used $\sigma_{1} \simeq 2 k A_{1}^{2}$. This
corresponds to the Friedmann's equation with a varying Newton's
constant $G_{N} \propto \sigma(\phi^{1})$. In other words, during
late cosmological times, solar system tests imply that our
universe's evolution must be described by the conventional
Friedmann's equation.

\subsection{Binary-pulsar tests}

\begin{figure}[ht]
\begin{center}
\includegraphics[width=0.45\textwidth]{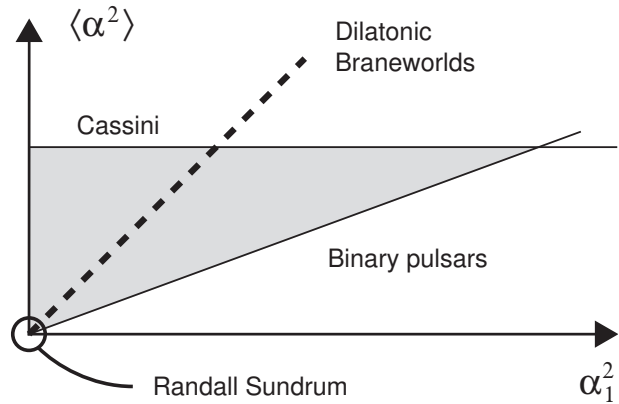}
\caption[Behavior of the fields]{The figure sketches the
constrained parameter space spanned by $\alpha_{1}^{2}$ and
$\langle \alpha^{2} \rangle$. The Randall-Sundrum model
corresponds to the point $(\alpha_{1}^{2} , \langle \alpha^{2}
\rangle)  = 0$, while dilatonic braneworlds are represented by the
curve $\langle \alpha^{2} \rangle = \alpha_{1}^{2}$.} \label{F4}
\end{center}
\end{figure}

Solar-system tests impose mild constraints on the $\beta$ parameter
and, consequently, on $\alpha^{2}_{1}$. For instance, bound
(\ref{S5: LLR}) gives $\alpha_{1}^{2} < 3.9 \times 10^{-3}$, which
is 3 orders of magnitude weaker than (\ref{S5: alpha}). It is
possible, however, to improve the bounds on $\alpha^{2}_{1}$ by
considering binary-pulsar tests; general relativity has shown to be
very successful in describing the physics of binary-pulsars. For
example, in the case of the Hulse-Taylor binary-pulsar PSR B1913+16,
three relativistic parameters have been determined with great
accuracy \cite{pulsar-exp}: the Einstein time delay parameter
$\gamma_{T}$, the periastron advance $\dot \omega$, and the rate of
change of the orbital period $\dot P$. It is well known that the
strong field conditions offered by binary-pulsars make the moduli to
behave nonperturbatively around them \cite{Pulsar1, Pulsar2},
modifying the predictions on $\gamma_{T}$, $\dot \omega$, and $\dot
P$ made by GR.  This offers strong constraints on $\beta$ and
$\gamma$; in particular, in the region $\beta < 0$ the following
result holds \cite{strong-field}
\begin{eqnarray}
\frac{\beta - 1}{\gamma - 1}  < 1.1.
\end{eqnarray}
Then, using Eqs. (\ref{S4: brane-edd3}) and (\ref{S4: brane-edd4}),
we obtain
\begin{eqnarray}
\alpha_{1}^{2} < 13.2 \, \langle \alpha^{2} \rangle, \label{S4: bound2}
\end{eqnarray}
which is a much stronger bound on the brane parameter
$\alpha_{1}^{2}$. Putting together bounds (\ref{S4: bound2}) and
(\ref{S5: alpha}) we can restrict the parameter space for braneworld
models considerably. Fig. \ref{F4} summarizes the discussed bounds
on the parameters of the theory.

\subsection{Geometrical interpretation of the constraints} \label{geom}

We now would like to extract some meaningful information out of the
constrained parameters $\alpha_{1}^{2}$ and $\langle \alpha^{2}
\rangle$. Let us start by observing that $\langle \alpha^{2}
\rangle$ measures the way in which $\sigma$ varies as a function of
the warp factor $\omega$. If the $\Sigma_{1}$ brane is located in a
strongly warped region, where $\omega$ varies quickly in terms of
$z$ while $\sigma$ does it slowly (as in the Randall-Sundrum model),
then we should generally expect $\langle \alpha^{2} \rangle$ to be a
small parameter. In contrast, if $\Sigma_{1}$ is in a region where
$\sigma$ varies importantly in terms of $z$ but $\omega$ does not,
then, in general, $\langle \alpha^{2} \rangle$ can be expected to
acquire arbitrary large values. To be more precise, let us
parameterize $\sigma$ as a function of $\omega$ (which has to be a
monotonically decreasing function of $\omega$), and consider the
profile shown in Fig. \ref{F5}:
\begin{figure}[ht]
\begin{center}
\includegraphics[width=0.45\textwidth]{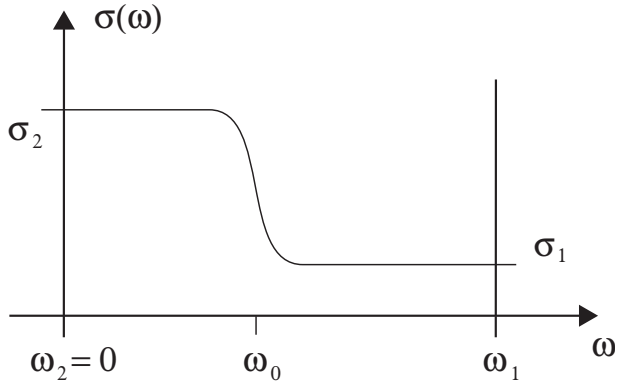}
\caption[Profile of sigma]{The figure shows a generic potential
$\sigma$ as a function of $\omega$. The parameter $\omega_{0}$ can
be adjusted to analyze different cases near the first brane (where
$\omega= \omega_{1}$).} \label{F5}
\end{center}
\end{figure}
The function $\sigma$ varies slowly in the region $\omega_{2} <
\omega < \omega_{0}$, with a value $\sigma \sim \sigma_{2}$. Then,
it has a steep variation around the value $\omega = \omega_{0}$.
And finally, it returns to a slowly varying function in the region
$\omega_{0} < \omega < \omega_{1}$, with $\sigma \sim \sigma_{1}$.
Additionally, assume that $\sigma_{2} \gg \sigma_{1}$ which is
generally the case if $\omega_{2} = 0$. Then $K^{2}$ and $W^{2}$
can be approximated to the following expressions:
\begin{eqnarray}
K^{2} &\simeq& 2 \frac{\omega_{0}^{2}}{\sigma_{1}}, \\
W^{2} &\simeq& 2 \frac{\omega_{1}^{2} - \omega_{0}^{2}}{\sigma_{1}}
+ 2 \frac{\omega_{0}^{2}}{\sigma_{2}}.
\end{eqnarray}
Thus if $\omega_{1} \gg \omega_{0}$, that is, if the geometry around
the brane is strongly warped, then we find
\begin{eqnarray} \label{stron-warp}
\langle \alpha^{2} \rangle \sim \frac{1}{2}
\frac{\omega_{0}^{2}}{\omega_{1}^{2}} \ll 1,
\end{eqnarray}
otherwise if $\omega_{1} \sim \omega_{0}$, then we find
\begin{eqnarray}
\langle \alpha^{2} \rangle \sim \frac{1}{2}
\frac{\sigma_{2}}{\sigma_{1}} \gg 1.
\end{eqnarray}
This example shows two extreme cases concerning the geometry
around the first brane $\Sigma_{1}$. The constraint (\ref{S5:
alpha}) thus allows us to infer that the geometry around the
braneworld $\Sigma_{1}$ is highly warped to agree with
observations.

\begin{figure}[ht]
\begin{center}
\includegraphics[width=0.45\textwidth]{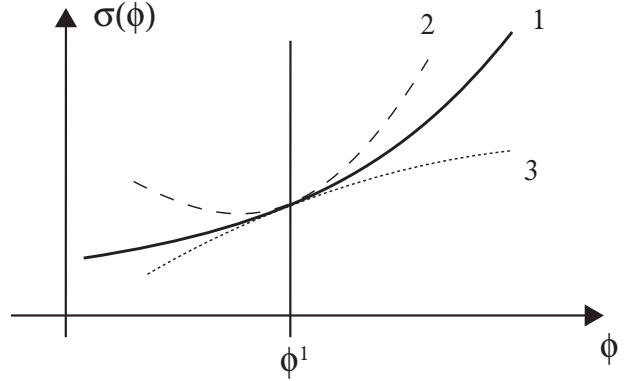}
\caption[Profile of sigma-phi]{The figure shows different
behaviors for $\sigma(\phi)$ about the first brane position
$\phi^{1}$. The curves are: Curve 1,
$\alpha_{1}^{2} = \langle \alpha^{2} \rangle$ (dilatonic
braneworlds); Curve 2, $\alpha_{1}^{2} < \langle \alpha^{2}
\rangle$; Curve 3, $\alpha_{1}^{2} > \langle \alpha^{2} \rangle$.}
\label{F6}
\end{center}
\end{figure}

We can also ask about the behavior of $\sigma$ around the brane. In
this case, the comparison between $\alpha_{1}^{2}$ and $\langle
\alpha^{2} \rangle$ tells us the growth of $\sigma$ around the
brane. For example, if $\alpha_{1}^{2} = \langle \alpha^{2}
\rangle$, the potential corresponds to the dilatonic case $\sigma
\propto \exp[\alpha \phi]$ (with $\alpha$ a constant), whereas if
$\alpha_{1}^{2} < \langle \alpha^{2} \rangle$, the tendency of
$\sigma$ is to grow faster than the dilatonic case, and if
$\alpha_{1}^{2} > \langle \alpha^{2} \rangle$, then the tendency is
to grow slower. Fig. \ref{F6} shows the different cases.
At present, bound (\ref{S4: bound2}) is unable to provide evidence
in favor on any of these cases.

\subsection{Can we live in the second brane?}

Up to now we have considered the case of tests being performed at
the first brane. We can repeat our analysis and compute the
Eddington's parameters for the second brane $\Sigma_{2}$. In this
case it is found:
\begin{eqnarray}
\eta^{2} = \frac{1}{3} \left[ 1 + \frac{\sigma_{2}}{2 k A_{2}^{2}}
\right],
\end{eqnarray}
which, after considering the cosmological evolution of the branes
provides the result: $\gamma - 1 \simeq - 2$. We therefore
conclude that our universe cannot be localized to the second
brane.

\section{Conclusions} \label{S5}

In this paper we have considered the low energy regime of 5D
braneworld models with a bulk scalar field. The present setup (BPS
braneworlds) included the Randall-Sundrum model and dilatonic
braneworlds as particular cases [different models are distinguished
by the brane potential $\sigma(\phi)$]. We discussed the way in
which solar system and binary pulsar tests can be used to constrain
the bulk geometry and the configuration of the branes. For this, it
was first necessary to understand the cosmological behavior of the
branes; we found that the negative tension brane is generally
expected to evolve towards a null warp-factor state, meaning that at
late cosmological times both branes end up interacting weakly.

At late cosmological times the positive tension brane is well
described by a single scalar-tensor theory of gravity [see Eq.
(\ref{onebrane})]. Then, having defined the parameter $\alpha =
\sigma^{-1} \partial_{\phi} \sigma$, it was found that the
post-Newtonian Eddington parameters $\beta$ and $\gamma$ can be
expressed in terms of two brane parameters, $\alpha_{1} =
\alpha(\phi^{1})$ and $\langle \alpha^{2} \rangle$, containing
information of the bulk geometry near the first brane. In terms of
these parameters, dilatonic braneworlds corresponds to the case:
$\langle \alpha^{2} \rangle = \alpha_{1}^{2} =\,$const.; whereas the
Randall-Sundrum model corresponds to: $\langle \alpha^{2} \rangle =
\alpha_{1}^{2} = 0$. We found two constraints: $\langle \alpha^{2}
\rangle < 1.5 \times 10^{-6}$ and $\alpha_{1}^{2} < 13.2 \, \langle
\alpha^{2} \rangle$, which allowed us to infer that the geometry
around the braneworld is strongly warped [see Eq.
(\ref{stron-warp})].

The analysis made in this paper should allow a better understanding
of generic 5D braneworld models and their low energy phenomenology.
The present results can be used together with other tests to further
constrain the role of $\phi$ (the bulk scalar field), as well as
other moduli, in the 4D standard model of physics. Examples of this
are the possibility of $\phi$ being a source for inflation
\cite{infl}, the variation of constants \cite{variconst}, a
chameleon field \cite{chameleon}, or dark energy. In the later case,
it would be interesting to explore the reconstruction of the shape
of $\sigma(\phi)$ as a function of $\phi$ out of measurements on the
dark energy evolution \cite{reconst1, reconst2}.

\section*{Acknowledgements}

The author is grateful to Anne C. Davis, Philippe Brax, and T.M.
Eubanks for useful comments and discussions. This work is supported
in part by DAMTP (Cambridge) and MIDEPLAN (Chile).

\appendix

\section{Einstein's frame sigma model} \label{A1}

Here we provide the Einstein's frame sigma-model connections
$\gamma_{c}^{a b}$. They are given by the expression
\begin{eqnarray}
\gamma_{c}^{a b} = \frac{1}{2} \gamma_{c d} \left[ \partial^{a}
\gamma^{d b} + \partial^{b} \gamma^{a d} - \partial^{d} \gamma^{a
b} \right],
\end{eqnarray}
where derivatives $\partial^{a}$ are made with respect to
$\omega_{a}$. The nonzero connections are
\begin{eqnarray}
\gamma^{1 1}_{1} &=& \frac{2}{\omega_{1}} \left[ \alpha_{1}^{2} -
\frac{2 k A_{1}^{2}}{\sigma_{1}} + \frac{2
k A_{1}^{2} \alpha_{1}^{2} W^{2}}{\sigma_{1} K^{2}} \right], \nonumber\\
\gamma^{2 2}_{2} &=& \frac{2}{\omega_{2} } \left[ \alpha_{2}^{2} +
\frac{2 k A_{2}^{2}}{\sigma_{2}} - \frac{2 k A_{2}^{2}
\alpha_{1}^{2} W^{2}}{\sigma_{2} K^{2}} \right], \nonumber\\
\gamma^{1 1}_{2} &=&  \frac{4 k \alpha_{1}^{2}
\omega_{2}}{\sigma_{1} K^{2}} , \qquad
\gamma^{2 2}_{1} =  - \frac{4 k \alpha_{2}^{2} \omega_{1}}{\sigma_{2} K^{2}} , \nonumber\\
\gamma^{1 2}_{1} &=&   \frac{2 k \omega_{2}}{\sigma_{2} W^{2}},
\qquad \gamma^{1 2}_{2} =  - \frac{2 k \omega_{1}}{\sigma_{1}
W^{2}}.
\end{eqnarray}

\section{2PN level parameters} \label{2PN}

At the second post-Newtonian (2PN) level (order $1/c^{4}$) it is
possible to introduce two new parameters $\epsilon$ and $\zeta$,
which play a similar role to $\beta$ and $\gamma$ \cite{2PN}. Here
we consider their definition for completeness. These parameters are
\begin{eqnarray}
\epsilon &=& \frac{ \beta^{a b c} \eta_{a} \eta_{b} \eta_{c} }{(1
+ \eta^{2})^{3}}, \label{S4: epsilon} \\
\zeta &=& \frac{ \beta^{a c} \beta_{c}^{b} \eta_{a} \eta_{b} }{(1
+ \eta^{2})^{3}}, \label{S4: zeta}
\end{eqnarray}
where $\beta^{a b c} = D^{a} D^{b} \eta^{c}$. A straightforward but
tedious calculation shows that
\begin{eqnarray}
\beta^{a b c} \eta_{a} \eta_{b} \eta_{c} &=& - \frac{6}{1 - 3 \eta^{2}} \left( \beta^{a b} \eta_{a} \eta_{b} \right)^{2} \nonumber\\
&& - \frac{8}{9} \alpha_{1}^{2} \left( 1 - 3 \eta^{2} \right)^{2}
\left[ 1 - 2 \alpha_{1}^{2} \frac{W^{2}}{K^{2}} \right] \nonumber\\
&& - \frac{16}{27} \alpha_{1}^{2} \left( 1 - 3 \eta^{2} \right)^{3} \frac{\partial \alpha_{1}}{\partial \phi^{1}} , \\
\beta^{a c} \beta_{c}^{b} \eta_{a} \eta_{b} &=& - \frac{1}{3}
\left( 1 - 3 \eta^{2} \right) \beta^{a b} \eta_{a}
\eta_{b} \nonumber\\
&& - \frac{2}{9} \alpha_{1}^{2} \left( 1 - 3 \eta^{2} \right)^{3}
\left[ 1 - 2 \alpha_{1}^{2} \frac{W^{2}}{K^{2}}  \right] \nonumber\\
&& + \frac{4}{27} \alpha_{1}^{4} \left( 1- 3 \eta^{2} \right)^{2}
\frac{\omega^{2}_{2} \sigma_{1}}{\omega^{2}_{1} \sigma_{2}}.
\end{eqnarray}
Using Eqs. (\ref{S4: brane-edd1}) and (\ref{S4: brane-edd2}), and
the fact that $\eta^{2} \simeq 0$, we find
\begin{eqnarray}
\epsilon &=& \frac{2}{3} (1 - \beta) + \frac{4}{27} \alpha_{1}^{2}
\frac{\omega^{2}_{2}
\sigma_{1}}{\omega^{2}_{1} \sigma_{2}} \nonumber\\
&& + \frac{2}{9} \frac{\alpha_{1}^{2}}{ \langle \alpha^{2} \rangle
} \left( \alpha_{1}^{2} -   \langle
\alpha^{2} \rangle  \right), \\
\zeta &=& - 24 (1 - \beta)^{2} - \frac{16}{27} \alpha_{1}^{2}
\frac{\partial \alpha_{1}}{\partial \phi} \nonumber\\
&& + \frac{8}{9} \frac{\alpha_{1}^{2}}{ \langle \alpha^{2} \rangle
} \left( \alpha_{1}^{2} -   \langle \alpha^{2} \rangle  \right).
\end{eqnarray}
In principle, these parameters can be used to further constrain 5D
braneworld models.

\end{document}